# "Shaking in 5 seconds!" A Voluntary Smartphone-based Earthquake Early Warning System

Rémy Bossu[1,2], Francesco Finazzi[3*], Robert Steed[1], Laure Fallou[1], István Bondár[4]

[1]European-Mediterranean Seismological Centre; Arpajon, France.

[2]CEA, DAM, DIF, 91297; Arpajon, France.

[3]University of Bergamo, Department of Management, Information and Production Engineering; Dalmine, Italy.

[4] ELKH Research Centre for Astronomy and Earth Sciences, Geodetic and Geophysical Institute; Budapest, Hungary.

*Corresponding author. Email: francesco.finazzi@unibg.it

## Abstract

Public earthquake early warning systems have the potential to reduce individual risk by warning people of an incoming tremor but their development has been hampered by costly infrastructure. Furthermore, users' understanding of such a service and their reactions to warnings remains poorly studied. The smartphone app of the Earthquake Network initiative turns users' smartphones into motion detectors and provides the first example of purely smartphone-based earthquake early warnings, without the need for dedicated seismic station infrastructure and operating in multiple countries. We demonstrate here that early warnings have been emitted in multiple countries even for damaging shaking levels and so this offers an alternative in the many regions unlikely to be covered by conventional early warning systems in the foreseeable future. We also show that although warnings are understood and appreciated by users, notably to get psychologically prepared, only a fraction take protective actions such as "drop, cover and hold".

## Introduction

Public earthquake early warning (PEEW) systems aim to warn people of imminent shaking through the rapid detection of earthquakes. They strive to reduce an individual's risk by allowing their users to take protective actions (such as "drop, cover and hold") in the seconds or tens of seconds separating the warning from ground shaking at the user's location. They were deployed first in 1991 in Mexico City (1) and then in Japan in 2007 (2). Despite this desirable goal and the existence of a number of other implementations, such as ShakeAlert® in the Western US (3, 4) and some private initiatives in Mexico and Chile, so far PEEW systems have not been put into service more widely, even in regions of high earthquake hazard, because they require dense, real time, and robust seismic and communication networks (5). Furthermore, PEEW evaluations have mainly focused on technical performance (e.g., rapidity, false/missed alert rates) with only a few studies carried out from users' perspectives to assess how the service is valued and, more importantly, whether users react or not after receiving the warning (1, 2) or how they anticipate to react for such a future service (6). This situation has led to a lack of actual assessment of PEEW in terms of individual risk reduction so that key parameters such as the public's tolerance to false and missed alerts remain unknown, making it difficult to develop informed and efficient warning strategies (7, 8).

Smartphones, due to their internal accelerometers, communication capabilities and their ubiquity were almost immediately identified for their low-cost potential for earthquake early warning (9, 10). If the feasibility of building a monitoring network from smartphones has been demonstrated, previously documented smartphone-based systems have not delivered their own early warnings (11, 12). On the other hand, the Earthquake Network (EQN) initiative (13, 14, 15) implements the first smartphone-based PEEW system that detects earthquakes in real time by turning users' smartphones into motion monitoring stations and also publishes the earthquake warnings that the network generates.

The resulting monitoring network is fully dynamic, with new users often joining after felt earthquakes and some leaving over time. EQN, available in 8 languages, has grown its userbase since its inception in 2012 with 6.5 million downloads and 650,000 active users in 2020 but this is its first actual evaluation in terms of early warning.

The EQN app installation turns participants' smartphones into real time seismic detectors by monitoring their internal accelerometers while their phones are charging. When an active (i.e., charging) smartphone senses an acceleration above a noise-dependent threshold a smartphone trigger is sent to the EQN servers and time stamped upon reception. No attempt is made to analyze seismic waveforms. A detection occurs when concurrent triggers within 30 km of each other exceed a dynamic threshold that is a function of the actual number of active smartphones and of the desired false alarm probability, a level currently set to one per year per country (14). A geo-located alert is issued at detection time to all users within 300 km of the detection location which is the average location of triggered smartphones and is taken as a proxy for the epicentral location. The alert is a smartphone notification with an easily recognizable sound and an automatic display of the epicentral location proxy, as well as a countdown in seconds to the estimated S-wave arrival time at the user location (Figure 1). Large earthquakes can cause several detections. To avoid multiple alerts for the same users only detections at least 300 km and 120 s apart are released.

The objectives of this work are to 1) evaluate EQN's detection performance, 2) demonstrate that it is capable of providing public earthquake early warning in multiple countries, and 3) assess the potential of EQN's contribution to individual risk reduction by studying EQN users' reactions after an actual early warning. Performance has been evaluated over a 26-month period (December 15$^{th}$, 2017 to January 31$^{st}$, 2020) during which the EQN data processing methodology was not modified. In addition, reaction to and understanding of early warning by EQN users has been inferred from an online survey of local EQN users in the felt area of the M8 2019 Peru earthquake.

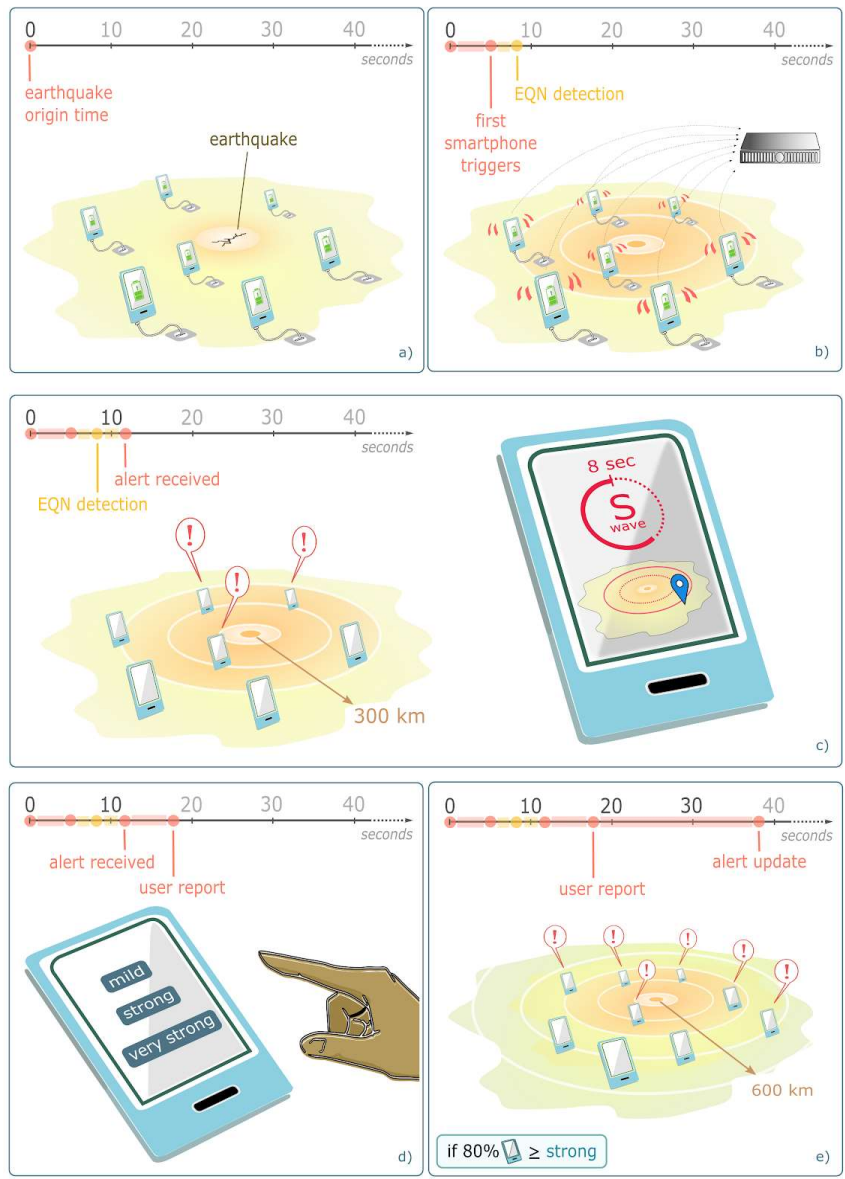

**Figure 1.** The EQN app turns a charging smartphone into a ground motion detector. Earthquakes are detected through a cluster of smartphone triggers. Once detected, an alert is issued to all users within a default distance of the detection and displays a countdown of the estimated S-wave arrival time at the user's location. Users can qualitatively report the level of shaking if they choose to, with 3 levels. This is intended to identify larger earthquakes as EQN does not provide magnitude estimates. If at least 10% of the users in the area of detection submit reports and 80% of these reports are "strong" or "very strong", a second alert is issued –typically 30 s after the first one– to users in an enlarged region (600 km by default). Users can opt in or out of the two alerts and customize alerting distances.

# Results

**EQN detection performance.** EQN detection performance in terms of latency, false detection rate and missed earthquake detections has been evaluated using 550 detections from the 3 countries (Chile, USA, and Italy) with at least 10 detections whose national catalogues possess both good location accuracy and coverage of low magnitude earthquakes, and where accelerometric data is available. Accurate locations are required to make proper estimates of the system's latency. Catalogues including low magnitude earthquakes are essential for both network sensitivity and false detection rate estimates as smartphone detections are possible down at least to M2 (11). Finally, accelerometric data was sought out from stations close to each detection location for a final consistency check against waveform data.

EQN detections were first associated in time and space with hypocenters from national catalogues, then among the potential candidates, an earthquake was considered as the source of the detection if the theoretical arrival time of the P-wave at the detection location was between 90 s before to 10 s after the detection time to allow for potential transmission delay and location uncertainties. This led to an initial association of 535 out of 550 detections. For this analysis, whenever an accelerometric station was available within 20 km of the detection location (410 out of 550 detections), the existence and time consistency of ground motion was visually checked. This inspection allowed the association of 4 additional detections. One was associated to a M3.8 earthquake at an unusually large distance of 350 km, and two to small magnitude earthquakes (M1.4 and M1.5 located 2 and 8 km from the detection) located through additional investigation by the Seismological Centre of the University of Chile. The fourth was found to be a secondary detection 800 km from epicenter of the March 1$^{st}$, 2019 Peru M7.0 earthquake. The false detection rate was ~2%.

The 539 associated detections are consistent with previous detectability studies of smartphone sensors (16). With half of them related to earthquakes below M4 (Figure 2), EQN detections also include events that are unlikely to generate strong shaking and therefore for which an early warning may not be necessary. However, comparison with independent data (Figure 2) indicates that nearly all EQN detections are likely to have also been felt, which make them relevant for rapid public information, even the few that were very low magnitude.

Assessment of EQN's rate of missed earthquakes is more complex than for traditional seismic monitoring networks as EQN geometry is governed by spatiotemporal variations in population distribution - higher in cities, lower in low population areas - and it constantly changes with app installations and deletions, and the number of active smartphones. Hence EQN detectability generally increases at night when more phones are charging. The rate of EQN earthquake detections was 3.1 times higher at night than during the day (Table 1). In Italy where the number of app users remained stable during the studied period (about 45,000) both earthquakes with M≥4.9 were detected as well as 4 out of 6 earthquakes with M≥4.5.

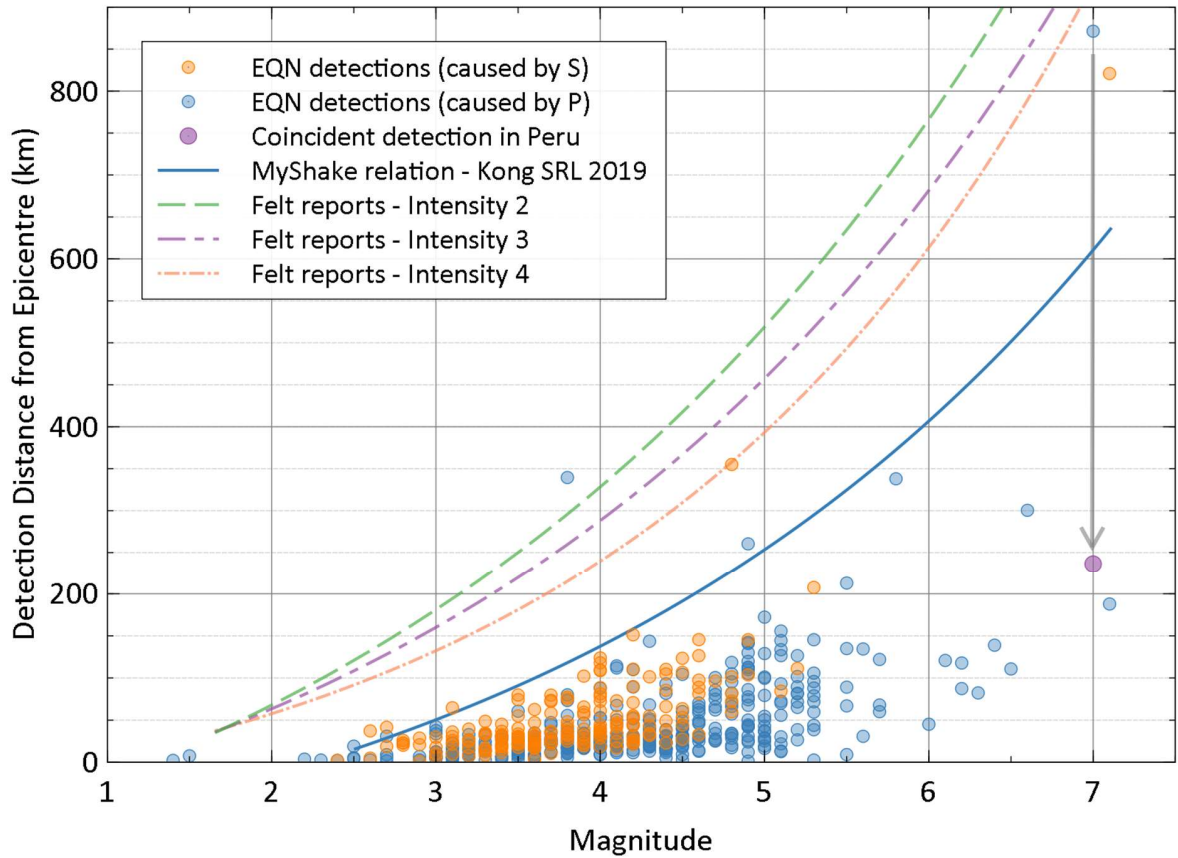

**Figure 2.** Distance between the location of the detection and the epicenter distance for 539 EQN associated detections as a function of magnitude. Blue and orange dots represent detections likely caused by P and S waves, respectively (the causative seismic phase is uncertain for epicentral distances below about 50 km). The M7 earthquake detected at more than 800 km epicentral distance was also detected in Peru, at about 250 km epicentral distance (arrow and purple dot). For comparison, the blue curve approximates the maximum distance to which smartphones operating MyShake app can detect earthquakes (16) while the 3 dashed lines approximate the 90% radial distance quantile of user-assigned intensities 2 (scarcely felt), 3 (weak) and 4 (largely observed) (based on the 1,528 global earthquakes between 2011 and end of October 2020 with at least 100 felt reports collected by the European-Mediterranean Seismological Center (EMSC).

| Country | Chile | USA | Italy | Total |
|---|---|---|---|---|
| **Detections** | 458 | 70 | 22 | 550 |
| **Detections associated with known earthquakes** | 449 | 70 | 20 | 539 |
| **Available accelerometric records** | 328 | 69 | 13 | 410 |
| **Magnitude (min; max)** | 1.4; 7.1 | 2.2; 7.1 | 2.4; 5.1 | 1.4; 7.1 |
| **Detection delay w.r.t. origin time (min; median; max in s)** | 4.8; 17.2; 209.0 | 4.3; 8.1; 42.5 | 3.4; 7.3; 11.0 | 3.4; 15.4; 209.0 |
| **Detection delay w.r.t. passing of triggering seismic wave (min; median; max in s)** | 0.5; 4.3; 12.1 | 2.0; 4.6; 10.2 | 1.8; 4.5; 5.9 | 0.5; 4.3; 12.1 |
| **False detection rate (%)** | 2.0 | 0.0 | 9.1 | 2.0 |
| **Nighttime/daytime ratio** | 2.7 | 11.3 | 8.0 | 3.1 |
| **Source of catalogue** | CSN | USGS | INGV | |

**Table 1.** Summary statistics of detections. Associated detections are the number of EQN detections for which it was possible to identify the causative earthquake. The accelerometric record column gives the number of detections for which accelerometric data is available within 20 km of the detection location. Detection delays were computed with respect to the earthquake origin time and the most likely causative seismic phase. False detection rate is the ratio between the number of false detections and the total number of detections while the nighttime/daytime ratio is computed considering that day (7:00 a.m. - 10:59 p.m.) lasts twice the night. CSN: Centro Sismologico Nacional, Chile. INGV: Istituto Nazionale Geologia e Vulcanologia, Italy.

**Latency of earthquake detections from a dynamic monitoring network.** The shortest earthquake detection latencies, i.e., the time difference between earthquake origin time and alert issuance, are achieved when the hypocenter is close to regions where the EQN app is popular. This explains why the median detection time was around 7-8 s in Italy and USA, where all detected earthquakes were onshore and at crustal depth (<40 km) compared to 17 s in Chile where a significant proportion of detected earthquakes were offshore and/or at intermediate depth (Table 1).

A limited comparison of earthquake detection times can be performed with ShakeAlert®, the operational EEW system which aims to cover the West Coast of the USA with 1,700 seismic stations (3, 4). Four earthquakes, the M7.1 Ridgecrest mainshock and 3 of its aftershocks ranging in magnitude from 3.8 to 4.5 were detected by both systems. Excluding the case of the mainshock (discussed below), EQN latencies are larger by an average 1.6 s (7.6 s versus 6.0 s averages for EQN and ShakeAlert® respectively) which is rather small considering the difference in technology levels. The Ridgecrest sequence exemplifies how EQN performance can rapidly change due to sudden app adoption. This sequence started with a M6.4 foreshock 36 hours before the mainshock. The foreshock was not detected due to a lack of EQN users in California at the time. However, this foreshock led to EQN installations in sufficient number in the Los Angeles (LA) area (but not in the epicentral region) so that the mainshock was detected in LA, 200 km to the south of its epicenter. Seismic wave propagation times from epicenter to LA where it was detected explains the unusually large detection latency of 40 s (see Table 2). In turn, the mainshock led to new EQN installations at shorter epicentral distances leading to a drop of EQN detection latency to 8 s (median times) for the 27 subsequent detected M2.7 to M4.6 aftershocks (see Table 2).

| Magnitude | Origin time | ShakeAlert® detection delay (s) | EQN detection delay (s) | EQN detection distance (km) |
|---|---|---|---|---|
| 3.9 | December 12$^{th}$, 2019 08:24:32.6 | 6.8 | 10.4 | 20 |
| 3.8 | December 5$^{th}$, 2019 08:55:31.65 | 5.7 | 5.4 | 10 |
| 4.5 | October 15$^{th}$, 2019 05:33:42.81 | 5.6 | 7.2 | 3 |
| 7.1 | July 6$^{th}$, 2019 03:19:53.04 | 6.9 | 40.0 | 188 |

**Table 2**. Detection latencies for the 4 earthquakes detected by both ShakeAlert® and EQN. These 4 earthquakes were detected in California and they followed the M7.1 Ridgecrest mainshock. ShakeAlert® detection times were retrieved from Chung et al. (17) for the M7.1 Ridgecrest earthquake in California and from http://earthquake.usgs.gov for the others.

To evaluate EQN's intrinsic latency, the wave propagation time of the most probable causative seismic phase from the epicenter to the EQN's detection location is subtracted from alert issuance latency. Note that this is an overestimation of cumulative processing and transmission delays as it implicitly assumes that acceleration (i.e., the monitored parameter) peaks at seismic phase onset. This implies that the minimum and median latencies (0.5 s and 4.3 s respectively, Table 1) characterize the best detection latencies that the EQN system can offer. Such fast detection is an achievement considering EQN's low investment cost.

In summary, EQN detection latency with respect to origin times for crustal earthquakes in regions with a significant app audience is comparable (5-8 s) to latencies observed in systems such as ShakeAlert® and, in the best-case scenario, it could be as low as a couple of seconds.

**EQN warning times.** Warning time is defined for a given target intensity as the time delay between the issuance of the alert and when the S-wave arrives at the users' locations who experience that target intensity. Being computed for the slower and stronger S-wave, it assumes that the P-wave is imperceptible and that from a user point of view this is the delay between the alert issuance and the perceived tremor. It assumes that the maximum intensity is generated by the onset of S-wave. Warning times have been computed at target intensities 4 (largely observed), 5 (strong) or 6 (slightly damaging) for all detected earthquakes (without any geographical restrictions) greater than M4.5 in Italy and USA and greater than M5 in the rest of the world. Intensities with respect to radial distance were estimated through intensity predictive equations (IPE) according to the validity domain of the considered IPE. Region-specific IPE have been used in the Western USA (18), and Italy (19) for crustal earthquakes (focal depth between 0 and 40 km). For all other regions, including deeper earthquakes, the same IPE (20) was used. This earthquake dataset being global, for the sake of homogeneity, earthquake parameters are from the US Geological Survey (USGS).

According to these estimations, over the 72 detected earthquakes greater than M4.5 or M5, EQN issued early warnings for target intensity 4 for 53 (74%) earthquakes of them (i.e., on average twice a month) located in 11 countries in North, Central and South America, Europe, and Asia (Figures 3 and 4). Among these, 18 events also benefited from a warning for target intensity 5 and for two earthquakes there was a warning for target intensity 6: M6.4 November 26th, 2019 Albania and M6.2 July 26th, 2019 Panama. As expected, for a given target intensity, warning times increase with increasing magnitude and for a given earthquake, they decrease with increasing target intensities. For earthquakes greater than M6, estimated warning times are typically more than 10 s for target intensity 4 and more than 5 s for target intensity 5 (see Figure 3), long enough for the user to take protective measures.

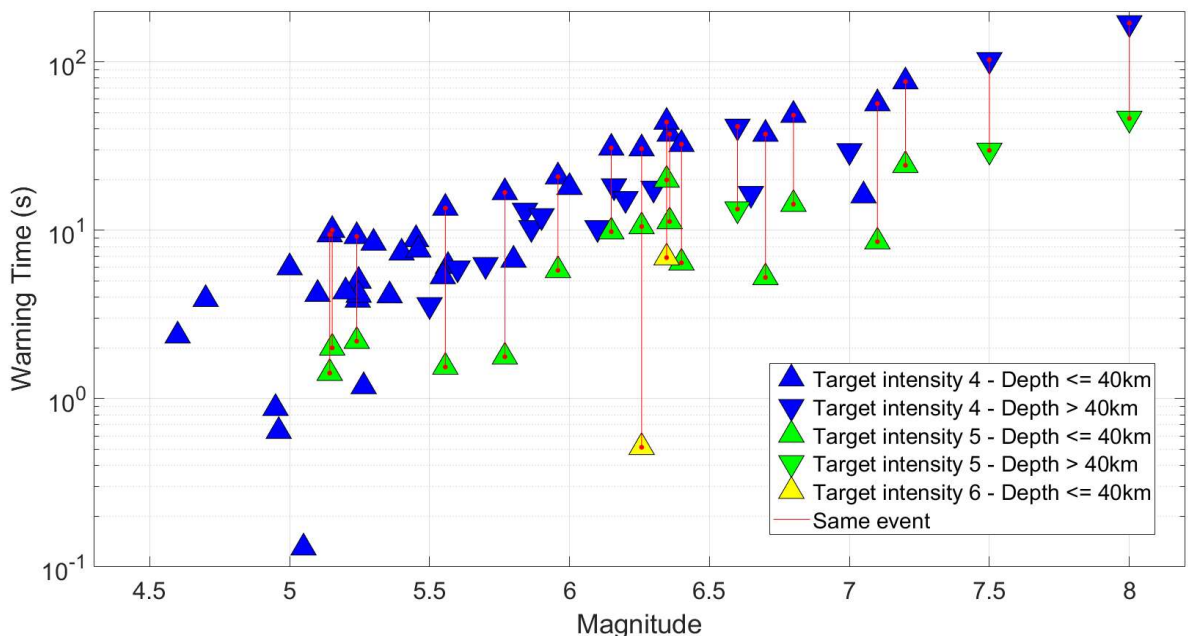

**Figure 3.** Estimated warning times for the 53 earthquakes detected worldwide with magnitude equal or greater than 4.5 with positive warning time. Blue, green, and yellow triangles depict warning times for target intensities 4, 5, and 6, respectively. Crustal and deep earthquakes are shown by triangles and inverted triangles, respectively. Warning times related to the same event are connected by red lines. For sake of clarity, magnitude is altered by a random shift of +/-(0.03, 0.06) for earthquakes sharing the same magnitude.

The warning time for target intensity 6 for the Panama earthquake is too short for individual protective action. However, for the Albania earthquake, which struck at night and killed 51 people, a warning time of 6.9 s for intensity 6 is estimated through the IPE, for a detection delay of 5.1 s after its occurrence, and a location of the detection 20 km from its epicenter.

According to the IPE, the isoseismal for intensity 6 was at 34 km from epicenter compared to 45 km from the empirical intensity-distance curve derived from about 4,000 eyewitnesses' reports crowdsourced for this event (21). This implies that the warning times derived from the IPE is likely underestimated by about 2 s for intensity 6 leading to a warning time for "slightly damaging" shaking exceeding 8 s. Based on the spatial distribution of EQN users at the time of the earthquake, and neglecting the transmission delay of the alert, we estimate that 1,005 of them received the early warning for intensity 6, 231 for intensity 5 and 632 for intensity 4. With approximately 800,000 inhabitants within 40 km of the epicenter, the proportion of warned individuals remains small in this case. Still, it proves that EQN can offer significant warning time for damaging shaking levels and so has the potential to lower individual seismic risk for its users.

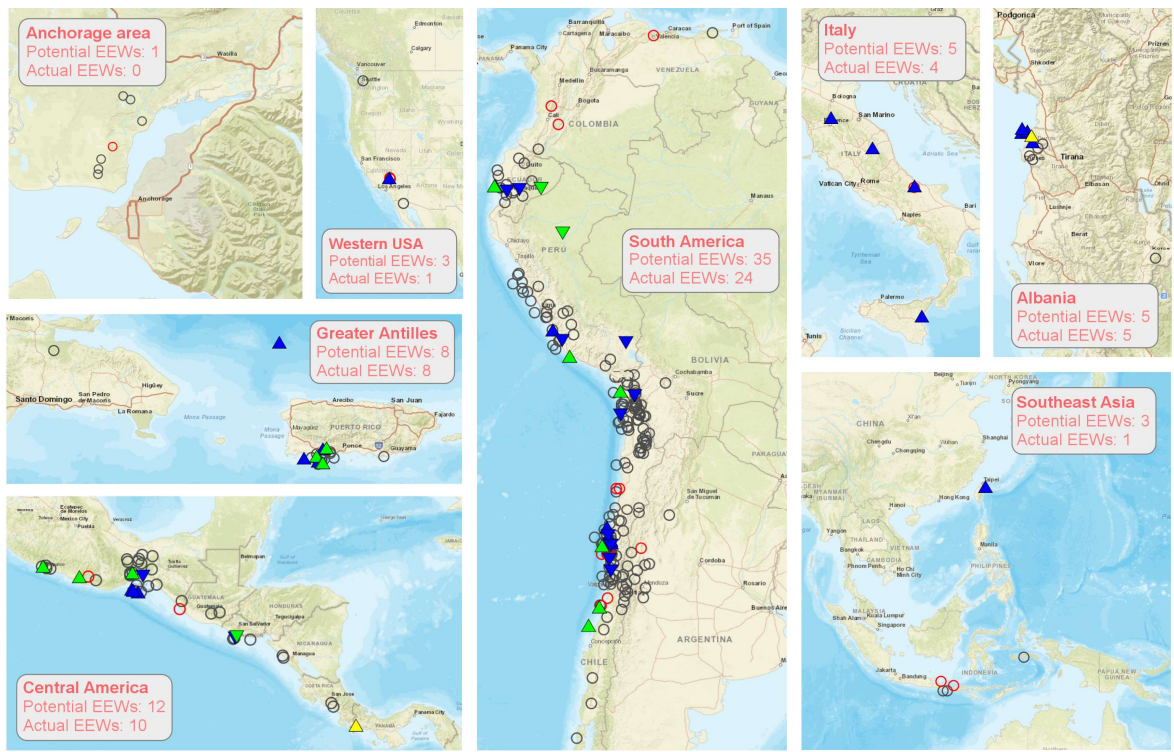

**Figure 4.** Geographical distribution of the 53 earthquakes for which a positive warning time is determined, shown as triangles (see Figure 3 for legend). All other EQN detected earthquakes of magnitude M4.5 or above are represented by circles, in red when the maximum onshore intensity

reached or exceeded intensity 4 (for which an EEW is theoretically possible) and in grey otherwise. The number of EEW in the legends indicates the number of positive warning times at intensity 4.

**Do EQN users take protective actions after a warning?** The reaction to, and understanding of, early warning has been assessed from an online survey of EQN users in the felt area of the M8 2019 Peru earthquake to evaluate EQN efficiency in terms of individual risk reduction. This earthquake had a focal depth of 120 km and generated two EQN detections, one in Peru and one in Ecuador. Alerts were issued for 599 users for intensity 5 and 54,228 for intensity 4, respectively. The survey was carried out from July 23$^{rd}$ to August 19$^{th}$, 2019. It was initiated through a message sent for technical reasons to all Spanish language EQN app, linking to an online questionnaire in Spanish. There were 61,863 users within 1,500 km of the epicenter, a distance where USGS and EMSC estimate the intensity between 3 and 4. 2,719 self-selected participants responded to the questionnaire; ⅔ of them declared to be between 500 to 1,000 km from the epicenter at the time of the earthquake, a range containing the capital cities of Quito and Lima. Most respondents (82%) declared previous earthquake experiences and 25% answered that they had experienced an EQN earthquake early warning before. 72% were convinced or strongly convinced of the usefulness of this app. Among these 2,719 self-selected respondents, 1,704 had the app at the time of the earthquake, while the others installed it following the earthquake. This first group described various experiences: 34% received EQN notification before feeling the shaking as expected from a PEEW system, 34% received it after having felt the shaking, 11% received the notification but did not feel the quake, 14% did not receive the notification while feeling the shaking, and 6% neither received the notification nor felt the quake.

Importantly, among the users who received the notification before feeling the shaking, 79% understood that a tremor was about to hit. This means they had a good comprehension of what an early warning is but when asked about their reaction (several answers possible), only 25% performed "drop, cover and hold", 10% ran outside, and 3% did nothing. Their priority was to warn relatives nearby (56%) or for the ones not in immediate proximity through social media (23%), and to wait for the shaking (36%).

This single study based on self-selected participants and on a single case shows that a low-cost smartphone based PEEW system can offer an actual early warning to some users even if the alert dissemination delay is unknown and may differ from one user to the next. However, in its current setting, and although the meaning of the notification is often understood, it only leads to adequate protective actions in a minority of cases, possibly because it does not answer an expressed priority need, which is to inform loved ones who may not have the app. The fact that EQN is appreciated by most of its users suggests that, despite EQN's inability to systematically guarantee an early warning, such a service combining early warning and rapid detection of felt earthquakes is valued by its users and constitutes a progress in public earthquake information.

## Discussion

The EQN initiative exploits smartphone ubiquity to create an operational network that provides an early warning service to its users. This service differs from conventional services in several aspects as EQN's alerting strategy is not based on predicted intensity and such predictions are not included in its warning messages. As a consequence, it also provides rapid information for small magnitude felt earthquakes for which no early warning is possible. This also removes errors due to differences between predicted and actual spatial distributions of shaking and bypasses the classic confusion among the public between magnitude and intensity. All of which simplifies the service's behavior and the content of its warning messages which otherwise can be difficult to understand from a user point of view, especially in a few seconds, as illustrated after the Ridgecrest earthquake where an alerting strategy had to be modified following users' feedback (8, 22). Also, although EQN users are volunteers and so EQN's alerting strategy has not yet been proven suitable for all audiences, receiving alerts for smaller earthquakes has not been identified by EQN's users as a weakness, and extending early warning service to also offer rapid public information for felt earthquakes seems generally to be an appreciated feature (22). Thus, EQN's early warning and rapid information services are a significant improvement for seismically active regions of the globe not yet covered by conventional PEEWs.

## Methods

**Datasets used in analysis.** Datasets were constructed from the events detected by the Earthquake Network (EQN) app between December $15^{th}$, 2017 and January $31^{st}$, 2020. This time range was chosen so that EQN's detection procedures would be stable during the entire period. There were 1,792 detections during this period in 19 countries. In order to perform quantitative analysis, 2 sub-datasets were extracted from this global dataset. These datasets are available as externally hosted supplementary material as Data S1 and Data S2.

**Data S1** is composed of 550 detections for examining the speed and location accuracy of EQN. Among the countries with a strong user base for the app, we chose to analyze the events in Chile, USA, and Italy due to the accuracy and completeness of their catalogues. Importantly, all three regions operate dense seismological station networks that are able to produce accurate event locations and magnitude estimates. An epicentral location inaccuracy of 15 km translates to a seismic phase arrival time change of 2-3 s which can become important in the case of EQN due to its rapid response times. All 3 regions also have dense accelerometer networks whose records were used to validate the EQN triggers. The USGS (USA) and INGV (Italy) catalogues of earthquake parameters were searched via FDSN requests while the CSN (Chile) catalogue was provided upon request. Calculations of the P and S seismic phases used the ak135 model and were carried out by the obspy Python library (see following sections for other calculation of other fields). The distributions of Data S1 with respect to magnitude and detection date are depicted in Figure 5.

**Data S2** was used for an analysis of EQN's early warning performance and consists of moderate to large magnitude earthquakes from around the world that were detected by EQN. This analysis employed intensity predictive equations (IPE) to estimate the intensities felt in regions that were warned of imminent shaking by the EQN app. The IPE equations' validities limited the analysis to earthquakes $\geq$ M5 in most of the world and $\geq$ M4.5 in Italy and USA (the equations are presented in the section **Calculation of Shaking Intensities**). The dataset is composed of 168 earthquakes and has 68 detections in common with Data S1. The main results from analysis of Data S2 can be seen in Figures 3 and 4. All of the earthquake parameters were obtained from the USGS catalogue for consistency.

There were also 3 earthquakes that were detected twice by EQN, normally such duplicate detections are suppressed automatically but all 3 earthquakes were large magnitude events (M7.0, M7.5 and M8.0) that led to EQN making detections at distances far from the epicenters. These 3 duplicate detections have been removed from the dataset for clarity.

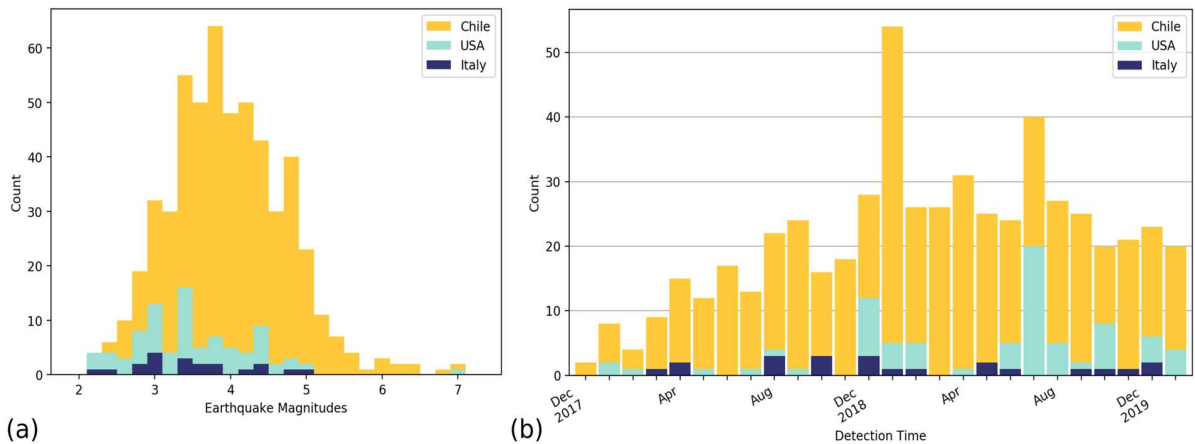

**Figure 5. (a)** This stacked Histogram shows that EQN detected earthquakes over a range of magnitudes in Chile, USA, and Italy. 539 out of the 550 EQN detections studied were associated with earthquakes with published parameters. **(b)** A stacked histogram of the number of EQN detections per month in Chile, USA, and Italy. A growth in the number of detections can be seen for Chile and the USA over this period.

**Association of Detections with Earthquakes.** For the purposes of the analysis, it is important to associate each EQN detection with earthquakes parameters held in an institute's catalogues of events. The following procedure was used for association:
1. Earthquakes were selected from the catalogue from 250 s before the time of the detection until 4 s afterwards.
2. Earthquakes were selected that are also within the association distance defined by each earthquake's magnitude (see Figure 6).
3. For each earthquake, the arrival time of the P waves at the EQN detection location was estimated using the ak135 model's speed of 8.04 km/s. The events whose P waves arrive within 90 s before the EQN detection and 10 s after the detection were chosen.
4. If multiple earthquakes remained in the selection, then the earthquake of the largest magnitude was chosen as the associated earthquake.

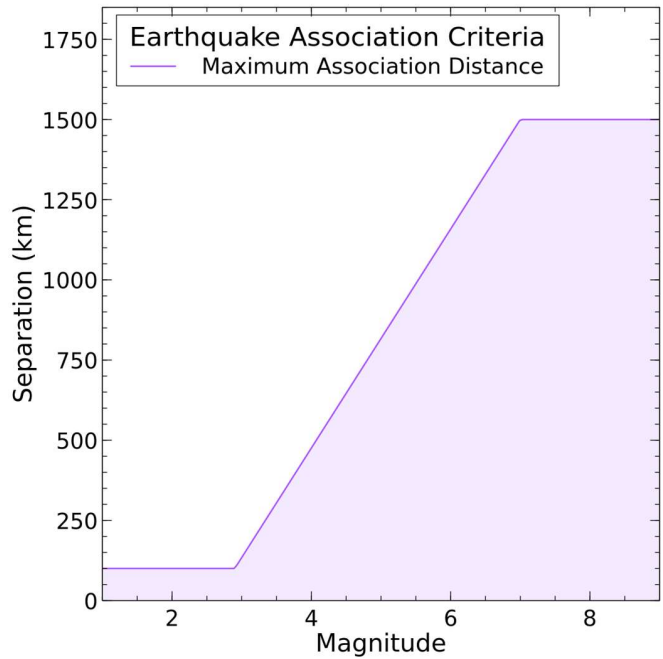

**Figure 6.** Association between an earthquake and EQN detection is allowed only if the separation between the epicenter and the EQN detection location is less than a threshold distance dependent upon the earthquake's magnitude as shown above.

**Causal Seismic Phase of EQN detections.** It has been found that EQN detections can be triggered by either P or S seismic phases (see Figure 7). The EQN detections were split heuristically into being caused by P or S phases using the criteria:

$$Caused\ by\ S\ if: (detection\ delay\ w.r.t.S > 0\ s)\ \&\ (detection\ delay\ w.r.t.P > 6s). \quad (1)$$

Note that distinguishing between P and S phases is less clear within 50 km of the epicenter since both arrive within a short interval of time. In addition, the EQN detections are triggered by strong motion due to the relative insensitivity of the smartphone accelerometers and the P/S phase arrival does not exactly coincide with the onset of motion strong enough to cause a detection.

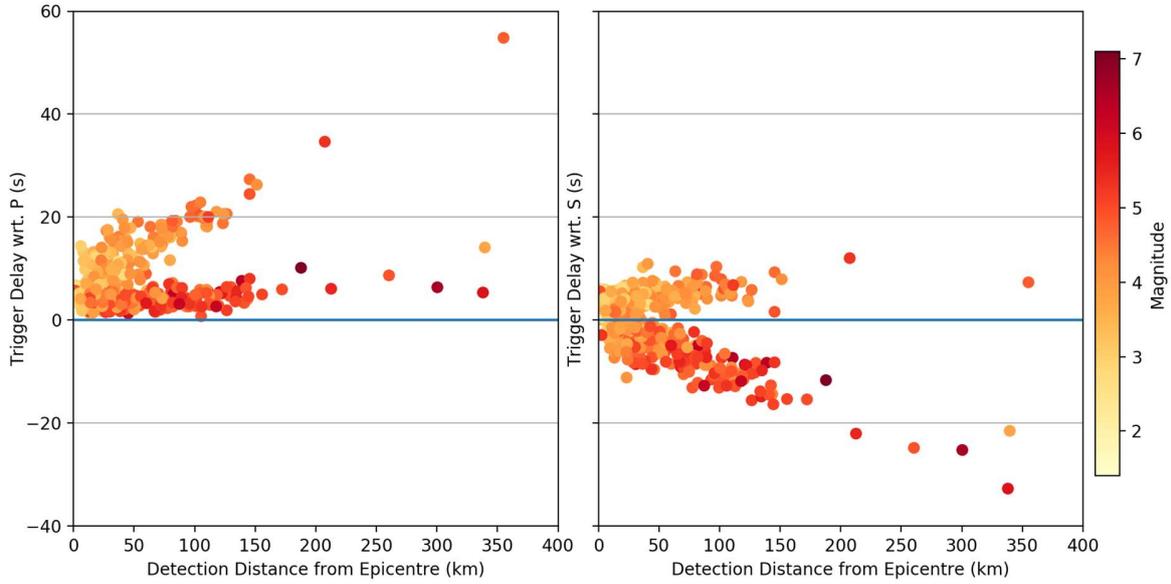

**Figure 7.** Determination of whether EQN detections follow the P or S seismic phase using Data S1. The arrival of the P and S phases at the detection location were calculated using the ak135 model and the latency between each phase arrival and the detection time is plotted against separation between the detection location and the epicenter. It can be seen that detections closely follow the passing of either the P or S phases and that EQN tends to detect larger magnitude earthquakes using the P wave.

**Calculation of Shaking Intensities.** Intensity predictive equations (IPE) were used to create the columns in the datasets (Data S1 and S2) and for the analysis of early warning times presented in the article. An IPE predicts the total felt intensity of shaking with respect to hypocentral distance for a given magnitude of earthquake. For a given delay from the origin time of an earthquake, the distance of the S phase from the epicenter can be calculated using the ak135 model and the intensity of shaking for this distance can then be calculated using the IPE. Alternatively, the distance at which the intensity reaches a certain value can be found and then the time at which the S phase passes this distance can be calculated in order to estimate whether there would be time for a warning to be given to people at this intensity.

To convert between epicentral and hypocentral distance the following equation was adopted:

$$r^2 = d^2 + 4R(R-d)\sin^2\left(\frac{s}{2R}\right), \quad (2)$$

where $r$ is the hypocentral distance, $d$ is the hypocentral depth, $R$ is the Earth's radius and $s$ is the epicentral distance.
For most earthquakes, the IPE from Allen et al. (2012) (20) were used, this formula is only valid for magnitudes >M5 and so we restricted the analysis accordingly:

$$\text{Intensity} = \begin{cases} 2.085 + 1.428M + 1.402\ln\sqrt{r^2 + R_m^2}, & r < 50\ km \\ 2.085 + 1.428M + 1.402\ln\sqrt{r^2 + R_m^2} + 0.078\ln\frac{r}{50}, & r \geq 50\ km \end{cases}, \quad (3)$$

where $M$ is the earthquake magnitude and:

$$R_m = -0.209 + 2.042\exp(M - 5). \quad (4)$$

For the Italian earthquakes, the IPE from Tosi et al. (2015) (19) was employed for crustal earthquakes (focal depth between 0 and 40 km):

$$\text{Intensity} = -2.15\log_{10}r + 1.03M + 2.31. \quad (5)$$

For the western USA, the IPE from Atkinson et al. 2014 (18) was used:

$$\text{Intensity} = 0.309 + 1.864M - 1.672\log_{10}\sqrt{r^2 + 14^2} - 0.00219\sqrt{r^2 + 14^2} + 1.77\max\left(0, \log_{10}\frac{r}{50}\right) - 0.383M\log_{10}\sqrt{r^2 + 14^2}. \quad (6)$$

**Comparisons with Strong Motion Waveforms.** For Data S1 (detections in Chile, USA, and Italy), a search was made using the FDSN protocol for accelerometer station waveforms within 20 km of each EQN detection. The waveforms were detrended, calibrated as acceleration measurements and bandpass filtered between 0.5-12 Hz. The waveform was also shifted in time to account for the difference in radial distance for the EQN detection location and the strong motion station with respect to the epicenter of the earthquake. The shift crudely assumed a seismic phase velocity of 8 km/s and the time shift was less than 1s in the majority of cases. The correction ensured that there was no confusion in causality for the analysis whereby the EQN detection occurred before the strong motion arrived.

Accelerometric data was found for 410 of the 550 detections in Data S1. The analysis demonstrated a strong correlation between strong motion and the EQN detections as would be expected and that it was also found that even small accelerations were able to cause EQN triggers (see Figure 8). The analysis also corroborated that the detections can be triggered by both P and S seismic phases (see also Figure 7 which shows this through a timing analysis) although it should be remembered that the strong motion necessary to cause triggers might follow a few seconds after the passing wavefront.

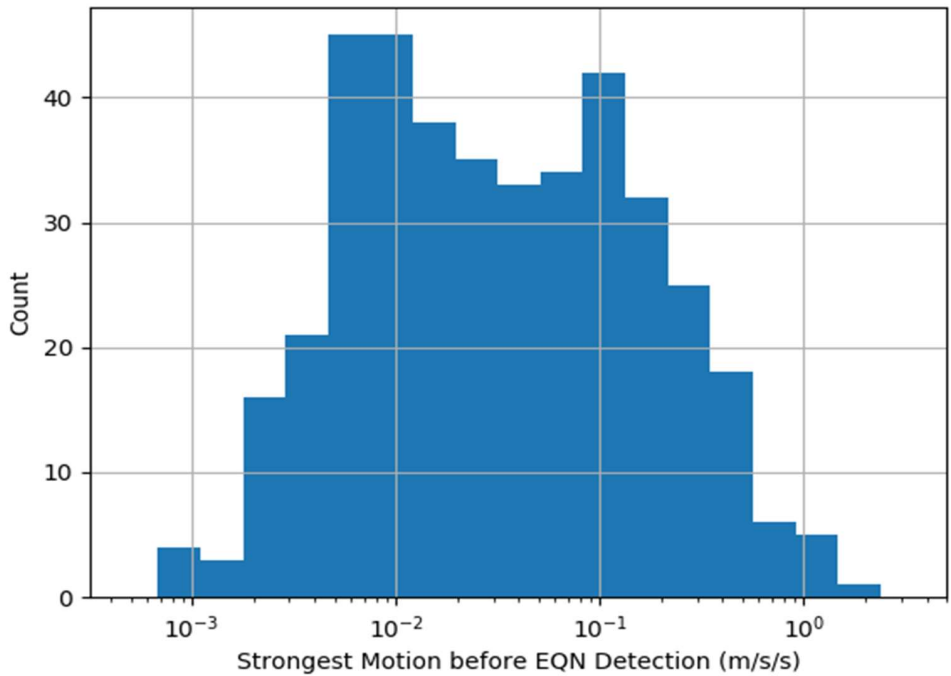

**Figure 8.** Histogram of the strongest acceleration found in the closest strong motion recording for each EQN detection in the 30 s period before detection. The results are only approximate since the level of shaking can significantly vary even over a distance of 10-20 km (23).

**References and Notes:**

1. Suárez, G., Novelo, D. & Mansilla, F. E. Performance evaluation of the seismic alert system (SAS) in Mexico City: a seismological and a social perspective. *Seism. Res. Lett*. **80**, 707–716 (2009). doi:10.1785/gssrl.80.5.707

2. Nakayachi, K., Becker, J. S., Potter, S. H. & Dixon, M. Residents' reactions to earthquake early warnings in Japan. *Risk anal.* **39**, 1723–1740 (2019). doi:10.1111/risa.13306

3. Kohler, M. D. et al. Earthquake early warning ShakeAlert system: west coast wide production prototype. *Seism. Res. Lett.* **89**, 99–107 (2018). doi:10.1785/0220170140

4. Given, D. D. et al. Revised technical implementation plan for the ShakeAlert system—An earthquake early warning system for the West Coast of the United States (No. 2018-1155). US Geological Survey. (2018). doi:10.3133/ofr20181155

5. Cremen, G. & Galasso, C. Earthquake early warning: recent advances and perspectives. *Earth-Sci. Rev*. **205**, 103184 (2020). doi:10.1016/j.earscirev.2020.103184

6. Becker, J. S., Potter, S. H., Vinnell, L. J., Nakayachi, K., McBride, S. K. & Johnston, D. M. Earthquake early warning in Aotearoa New Zealand: a survey of public perspectives to guide warning system development. *Humanit. Soc. Sci. Commun.* **7**, 1–12 (2020).

7. Allen, R. M. & Melgar, D. Earthquake early warning: advances, scientific challenges, and societal needs. *Annu. Rev. Earth Pl. Sc*. **47**, 361–388 (2019). doi:10.1146/annurev-earth-053018-060457

8. Cochran, E. S. & Husker, A. L. How low should we go when warning for earthquakes? *Science* **366**, 957–958 (2019). doi:10.1126/science.aaz6601

9. Minson, S. E. et al. Crowdsourced earthquake early warning. *Sci. Adv.* **1**, e1500036 (2015). doi:10.1126/sciadv.1500036

10. Kong, Q., Allen, R. M., Schreier, L. & Kwon, Y. W. MyShake: A smartphone seismic network for earthquake early warning and beyond. *Sci. Adv.* **2**, e1501055 (2016). doi:10.1126/sciadv.1501055

11. Kong, Q., Martin-Short, R., & Allen, R. M. Toward Global Earthquake Early Warning with the MyShake Smartphone Seismic Network, Part 1: Simulation Platform and Detection Algorithm. *Seism. Res. Lett.* **91**, 2206–2217 (2020). doi:10.1785/0220190177

12. Kong, Q., Martin-Short, R., & Allen, R. M. Toward Global Earthquake Early Warning with the MyShake Smartphone Seismic Network, Part 2: Understanding MyShake Performance around the World. *Seism. Res. Lett.* **91**, 2218–2233 (2020). doi:10.1785/0220190178

13. Finazzi. F. The earthquake network project: a platform for earthquake early warning, rapid impact assessment, and search and rescue. *Front. Earth Sci.* **8**, 243 (2020). doi:10.3389/feart.2020.00243

14. Finazzi, F. & Fassò, A. A statistical approach to crowdsourced smartphone-based earthquake early warning systems. *Stoch. Env. Res. Risk A*. **31**, 1649–1658 (2017). doi:10.1007/s00477-016-1240-8

15. Finazzi, F. The earthquake network project: Toward a crowdsourced smartphone-based earthquake early warning system. *Bull. Seism. Soc. Am.* **106**, 1088–1099 (2016). doi:10.1785/0120150354

16. Kong, Q., Patel, S., Inbal, A. & Allen, R. M. Assessing the sensitivity and accuracy of the MyShake smartphone seismic network to detect and characterize earthquakes. *Seism. Res. Lett*. **90**, 1937–1949 (2019). doi:10.1785/0220190097

17. Chung, A. I. et al. ShakeAlert earthquake early warning system performance during the 2019 Ridgecrest earthquake sequence. *Bull. Seism. Soc. Am.* **110**, 1904–1923 (2020). doi.org:10.1785/0120200032

18. Atkinson, G. M., Worden, C. B. & Wald, D. J. Intensity prediction equations for North America. *Bull. Seism. Soc. Am.* **104**, 3084–3093 (2014). doi:10.1785/0120140178

19. Tosi, P., Sbarra, P., De Rubeis, V. & Ferrari, C. Macroseismic intensity assessment method for web questionnaires. *Seism. Res. Lett*. **86**, 985–990 (2015). doi:10.1785/0220150127

20. Allen, T. I., Wald, D. J. & Worden, C. B. Intensity attenuation for active crustal regions. *J. Seismol*. **16**, 409–433 (2012). doi:10.1007/s10950-012-9278-7

21. Bossu, R. et al. Rapid public information and situational awareness after the november 26, 2019, Albania earthquake: lessons learned from the LastQuake system. *Front. Earth Sci*. **8**, 235 (2020). doi:10.3389/feart.2020.00235

22. Allen, R. M., Cochran, E. S., Huggins, T. J., Miles, S. & Otegui, D. Lessons from Mexico's earthquake early warning system. *Eos, Earth and Space Science News*. **99** (2018). doi:10.1029/2018EO105095

23. Ancheta, T. D., Stewart, J. P. & Abrahamson, N. A. Engineering characterization of earthquake ground motion coherency and amplitude variability. In: Effects of surface geology on seismic motion. 4th IASPEI/IAEEInternational Symposium, University of California Santa Barbara. 1–12 (2011).

**Acknowledgments:** The authors express their thanks to M. Corradini and M. Landès for fruitful discussions, H. Massone (CSN) for his identification of additional small magnitude earthquakes in Chile, as well as E. Calais and F. Cotton for their valuable suggestions for improvement of the manuscript. **Funding:** This article was partially funded by the European Union's Horizon 2020 Research and Innovation Program under grant agreement RISE No. 821115 and grant agreement TURNKey No. 821046. Opinions expressed in this article solely reflect the authors' views; the EU is not responsible for any use that may be made of information it contains. **Author contributions:** R.B.: conceptualization, supervision, methodology, validation and writing original draft. F.F.: conceptualization, resources, investigation, data curation, methodology, software, visualization, validation, review & editing. R.S.: investigation, data curation, methodology, software, visualization, validation, review & editing. L.F.: investigation, data curation, methodology, validation, review & editing. I.B.: data curation, validation, review & editing. **Competing interests:** Authors declare no competing interests.

**Data and materials availability**

Datasets analyzed in this article are available through GFZ Data Services at the following links.

Steed, R., Bossu, R., Finazzi, F., Bondár, I., & Fallou, L. (2021). Analysis of Detections by the Earthquake Network App between 2017-12-15 and 2020-01-31. V. 0.9. GFZ Data Services. https://doi.org/10.5880/fidgeo.2021.007 (Preview link: https://dataservices.gfz-potsdam.de/panmetaworks/review/9435b0cfcfb380571fe081844f6252c17b756a80a8344045609983874297bd69/)

R. Steed, R. Bossu, F. Finazzi, I. Bondár, L. Fallou. Analysis of Strong Motion Waveforms Near the Locations of Detections by the Earthquake Network App in Chile, the USA and Italy. V. 0.9. GFZ Data Services (2021). https://doi.org/10.5880/fidgeo.2021.002 (Preview link: https://dataservices.gfz-potsdam.de/panmetaworks/review/6aa55a072b38fd90bddb0ca01530ce273bc0c1967cdf84c12801a4fb8dfd070c/)

L. Fallou, R. Bossu, R. Steed, F. Finazzi, I. Bondár. A Questionnaire Survey of the Earthquake Network App's Users in Peru Following an M8 Earthquake in 2019. V. 0.9. GFZ Data Services (2021). https://doi.org/10.5880/fidgeo.2021.001 (Preview link: https://dataservices.gfz-potsdam.de/panmetaworks/review/12b6bb5153e39e7ed20dc0bda0d924dd39c60e0376617fb1ab2798cb0f1cb1a8/)